\documentclass[aps, prb, showpacs, twocolumn, amsmath, letterpaper]{revtex4-2}

\usepackage{graphics}
\usepackage{graphicx}

\usepackage{amssymb}
\usepackage{bm}

\usepackage[english]{babel}

\usepackage{amsmath}
\usepackage{graphicx}
\usepackage{wrapfig}

\usepackage[colorlinks=true, allcolors=blue]{hyperref}

\begin{document}
\title{Tuning the parity of superconductivity in CeRh$_2$As$_2$ via pressure}

\author{Hasan Siddiquee$^{1}$, Zackary Rehfuss$^{1}$, Christopher Broyles$^{1}$, Sheng Ran$^{1,2}$}
\affiliation{\\$^1$ Department of Physics, Washington University in St. Louis, St. Louis, MO 63130, USA
\\$^2$ The Institute of Materials Science and Engineering, Washington University in St. Louis, St. Louis, MO 63130, USA
}

\date{\today}

\begin{abstract}
In the recently discovered heavy fermion superconductor CeRh$_2$As$_2$, a magnetic field induced phase transition has been observed inside the superconducting state, which is proposed to be a transition from an even to an odd-parity superconducting state. The odd-parity superconducting state and its large upper critical field has been explained by local inversion symmetry breaking and consequent Rashba spin-orbit coupling. Here we report the experimental tuning of the parity of superconductivity in CeRh$_2$As$_2$ via applied pressure. Superconductivity is enhanced by pressure after its initial suppression, forming a second dome. However, the odd-parity state only exists in the first dome and the second dome is dominated by the even-parity state. This parity switch of superconductivity under pressure is consistent with the competition between Rashba spin-orbit coupling and interlayer hopping in determining the parity of superconductivity in systems with local inversion symmetry breaking. Our results provide insightful guidance on how to look for new odd-parity superconductors.


\end{abstract}
\maketitle
The discovery of odd-parity superconductors with spin-triplet pairing is a central goal of quantum materials research as they potentially host nontrivial topological excitations. Conventional wisdom has been to look for superconductivity in the vicinity of ferromagnetism. This route has led to the discovery of a collection of promising candidates, including the ferromagnetic superconductors UCoGe~\cite{huy2007}, URhGe~\cite{aoki2001}, UGe$_2$~\cite{saxena2000}, and the nearly ferromagnetic superconductor UTe$_2$~\cite{Ran2019}. Recently, a new avenue has opened with the discovery of the heavy fermion superconductor CeRh$_2$As$_2$ with critical temperature $T_c$ of 0.3~K, in which the odd-parity superconducting state is established by local inversion symmetry breaking and consequent Rashba-like spin-orbit coupling~\cite{khim2021field}.

A striking feature of the superconducting state of CeRh$_2$As$_2$ is the very high upper critical field $H_{c2}$~\cite{khim2021field}. When a magnetic field is applied along the $c$ axis, $H_{c2}$ reaches around 14~T at base temperature of 10~mK, far exceeding the Pauli limit of 0.6~T. Inside the superconducting state, a phase transition induced by applying a magnetic field along the $c$ axis was observed, which has been interpreted as a transition from a low field, even-parity, to a high field, odd-parity state. In addition, recent NMR measurements on CeRh$_2$As$_2$ have suggested an antiferromagnetic order within the superconducting state in zero magnetic field~\cite{Kibune2022,kitagawa2022} that disappears for magnetic field above 4~T applied along the $c$ axis. 

Theoretical studies suggest that local inversion symmetry breaking plays an essential role in establishing the odd-parity superconducting state in CeRh$_2$As$_2$~\cite{khim2021field,Cavanagh2022}. The tetragonal CaBe$_2$Ge$_2$-type structure of CeRh$_2$As$_2$ is globally centrosymmetric but breaks inversion symmetry locally at the Ce site enabling a Rashba spin-orbit coupling with an alternating sign on neighboring Ce layers. When a magnetic field is applied perpendicular to the Ce layers, the superconducting gap function on adjacent Ce layers related by inversion symmetry can have an opposite sign, leading to an odd-parity state. The $\vec{d}$ vector of the odd-parity state due to the Rashaba spin-orbit coupling is in the plane, causing the absence of the Pauli paramagnetic limit for $H \parallel c$. The angle dependence of the upper critical fields agrees well with the theoretical model, further confirming the superconductivity in higher magnetic fields is indeed the odd-parity state~\cite{Landaeta2022}.

This model originates from a theory developed in the context of layered superconductors with local inversion symmetry breaking that predicts a pair-density wave state~\cite{Fischer2011,Maruyama2012,Yoshida2013}. In principle it could apply to other quasi-2D superconductors with sublattice structure, However, the field-induced transition has not been widely observed. It turns out that the even to odd-parity transition is also suppressed by hopping between the two sublattices. In order for this transition to occur, the spin-orbit coupling needs to be larger than the inter-sublattice hopping, and it is not intuitively clear whether this condition can be realized in a bulk material. As a matter of fact, a relatively strong Rashba-like spin texture has been observed in a bilayer cuprate superconductor, but there is no evidence of a field-induced odd-parity state~\cite{Gotlieb2018}. To account for the observed transition in CeRh$_2$As$_2$, it has been proposed that, in addition to the local inversion symmetry breaking, the nonsymmorphic crystal structure allows the Rashba spin-orbit coupling to be larger than the interlayer hopping~\cite{Cavanagh2022}. 



The competition between the spin-orbit coupling and the interlayer hopping provides a possible route to experimentally switch the parity of the superconducting state. In this work we investigate the superconductivity of CeRh$_2$As$_2$ under applied pressure. From a simple tight binding point of view, interlayer hopping will increase as the lattice parameters decrease under pressure while spin-orbit coupling will remain the same, leading to a continuous decrease of the relative strength between spin-orbit coupling and interlayer hopping. The salient results of our study are the following: 1. after the initial suppression, superconductivity is enhanced by further increasing pressure, forming a double superconducting dome; 2. the odd-parity state is suppressed faster than the even-parity state, and only the even-parity state exists in the second superconducting dome. Experimental realization of such a parity switch of superconductivity under pressure not only provides strong support to the theoretical model, but also sheds light on how to look for new spin triplet superconductors.

\begin{figure}[t!]
    \includegraphics[width=1\linewidth]{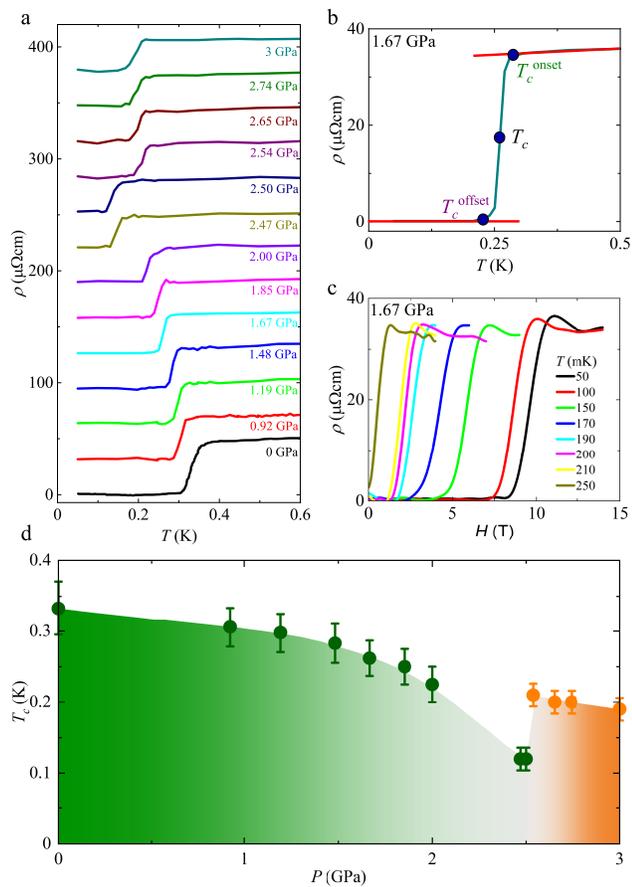}
    \caption{\textbf{(a)} Resistivity $\rho$ of CeRh$_2$As$_2$ as a function of temperature for different pressures. For clarity, the curves are shifted vertically by 33 $\mu$$\Omega$cm. \textbf{(b)} $\rho(T)$ data at 1.67~GPa as an example to illustration the criteria used to determine the $T_c$. $T_c$ is taken as the midpoint of resistivity drop upon the transition. The deviation from the zero resistivity is taken as the offset of $T_c$, while the deviation the normal state is taken as the onset of $T_c$. \textbf{(c)} Resistivity $\rho$ of CeRh$_2$As$_2$ as a function of magnetic field for different temperatures between 50 and 250~mK at 1.67~GPa. \textbf{(d)} $T_c - P$ phase diagram of CeRh$_2$As$_2$ showing a double superconducting dome.} 
\end{figure} 

Figure~1a shows the resistivity data as a function of temperature under various pressures up to 3.0~GPa. The superconducting transition is initially suppressed to lower temperatures upon increasing pressure, and then sharply enhanced to higher temperatures after 2.5~GPa, leading to a non-monotonic pressure dependence. The $T_c - P$ phase diagram constructed from $\rho(T)$ data is presented in Fig.~1d. $T_c$ is determined as the midpoint of the resistivity drop upon the transition, with the onset and offset of $T_c$ setting the error bar. The most important feature of the $T_c - P$ phase diagram is the double superconducting dome. The first dome has a maximum $T_c$ at ambient pressure and the second dome has a maximum $T_c$ at around 2.7~GPa. Close to ambient pressure, $T_c$ is suppressed at a very slow rate of -0.03~K/GPa. This is in sharp contrast to the estimation based on Ehrenfest relation using thermal expansion and specific heat measurements, which yield a suppression rate of -0.21~K/GPa~\cite{Hafner2022}, an order of magnitude higher than our experimental results. $T_c$ reaches a minimum value of 0.12~K at 2.5~GPa, and we did not find a pressure value where superconductivity is completely suppressed between the two domes.

\begin{figure*}[ht!]
    \includegraphics[width=1\linewidth]{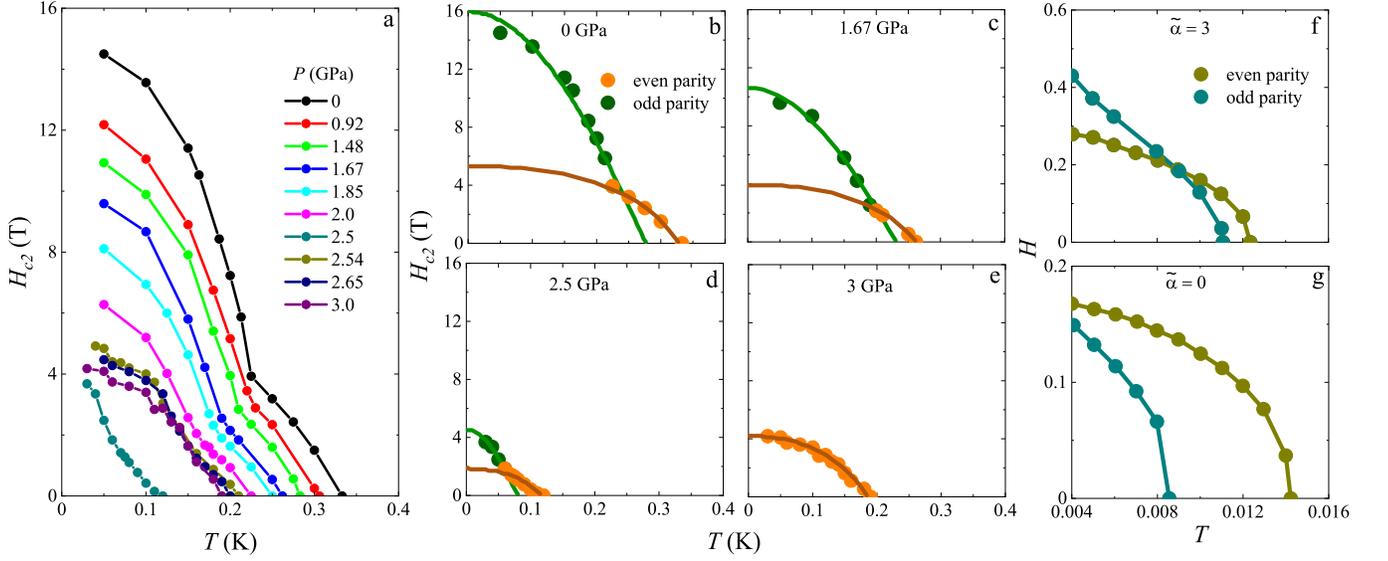}
    \caption{\textbf{(a)} $H_{c2}$ vs $T$ for different pressures. \textbf{(b-e)} Fits to the upper critical fields for even-parity (orange line) and odd-parity (green line) states for 0, 1.67, 2.5 and 3~GPa. Only even-parity state exists at 3~GPa. \textbf{(f-g)} Calculated $H-T$ diagrams for bilayer model with Rashba spin-orbit coupling for $\widetilde{\alpha}$ = 3 and 0~\cite{Nogaki2022}.} 
\end{figure*}

\begin{figure}[ht!]
    \includegraphics[width=1\linewidth]{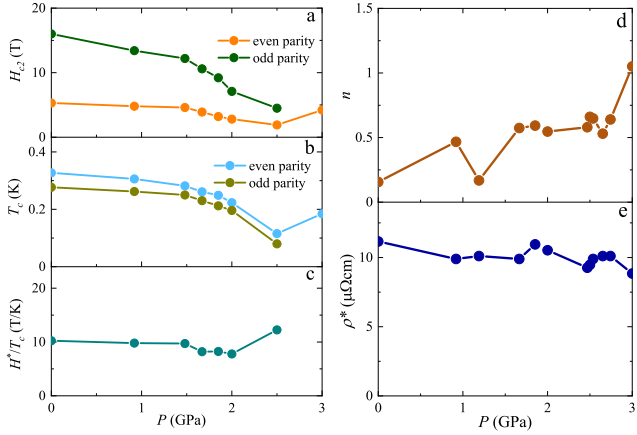}
    \caption{\textbf{(a)} $H_{c2}$ extracted from the fit as a function of $P$ for even and odd-parity superconducting state. \textbf{(b)} $T_{c}$ as a function of $P$ for even and odd-parity superconducting state. $T_{c}$ of the odd-parity state is extracted from the fit. \textbf{(c)} $H^\ast/T_c$ as a function of $P$, where $H^\ast$ is critical field for the even to odd transition and $T_c$ is the critical temperature of the even-parity state. \textbf{(d)} Power law expononet of the fit to the normal state as a function of $P$. \textbf{(e)} $\rho^{\ast} = \rho - \rho_0$ as a function of $P$ showing no obvious decrease.} 
\end{figure}

Having established the $T_c-P$ phase diagram, we now investigate the parity of the superconductivity inside the two domes. For this purpose, we measured resistivity as a function of magnetic field at various temperatures and extracted the upper critical field $H_{c2}$ for each pressure as shown in Figure~2. $\rho (H)$ shows hysteresis going through the superconducting transition and we use the data upon down sweep of magnetic field to extract $H_{c2}-T$ for all the pressures for consistency. At ambient pressure, when the magnetic field is applied along $c$ axis, there is a transition from even to odd-parity superconducting state, evidenced in ac magnetic susceptibility and specific heat measurements~\cite{khim2021field}. Resistivity does not show the transition as it is always zero inside the superconducting state. However, $H_{c2}-T$ shows a clear kink at $H^\ast$ separating the even and odd-parity states, which can be used to infer the even to odd state transition. Under pressure, the even to odd transition persists up to 2.5~GPa, and $H^\ast$ is suppressed to lower magnetic fields. Above 2.5~GPa, no obvious kink is observed indicating there is no longer a transition from even to odd-parity state. 

To investigate the nature of the superconductivity in more detail, we fit the $H_{c2}-T$ curve to the following expression that includes both Pauli paramagentic effect and orbital effect~\cite{khim2021field,Landaeta2022}: 
\begin{equation}
  Ln(t) = \int_{0}^{\infty}du \frac{[1-F+Fcos(\frac{Hgu}{H_Pt})]exp(\frac{-Hu^2}{\sqrt{2}H_{orb}t^2}) - 1}{sinhu},
\end{equation}
where $t = T/T_c$, $H_P$ and $H_{orb}$ are Pauli and orbital limit, and $F$ quantifies the pair breaking due to the Pauli paramagnetic effect ($F$ = 1 for even-parity state and $F$ = 0 for odd-parity state). The representative $H_{c2} - T$ curve and the fit for pressure below and above 2.5~GPa are shown in Fig.~2. At 1.67~GPa, the low field part can fit well to the expression for the even-parity state while the high field part can fit well to the expression for odd-parity state, similar to the ambient pressure results. As the pressure increases, the upper critical field of the odd-parity state is quickly suppressed, faster than that of the even-parity state, as shown in Fig.~3. At 2.5~GPa where $T_c$ is suppressed to a minimum value, we can still fit the low field part to the even-parity state and the high field part to the odd-parity state. On the other hand, at 3~GPa, $H_{c2}-T$ in the whole field range fits well to the expression for the even-parity state, with a Pauli critical field that is enhanced due to the spin-orbit coupling compared with the Clogston-Chandrasekhar limit of $\sim$ 0.3~T. This leads to the main finding of this work: the superconductivity in the second dome is dominated by even-parity state. 

What is the mechanism to cause the parity change of the superconducting state? At ambient pressure, the odd-parity state is also quickly suppressed when the magnetic field is turned away from the $c$ axis~\cite{Landaeta2022}. This has been explained in terms of the in plane $\vec{d}$ vector of the odd-parity superconductivity which leads to an absence of Pauli limiting for $H \parallel c$ and the presence of Pauli limiting for $H \parallel ab$. In the current study, the magnetic field is always applied along the $c$ axis. The suppression of odd-parity state must be due to a different mechanism. Note that in the context of layered superconductors with local inversion symmetry breaking, the amplitude of the odd-parity state is proportional to $\alpha_R$/$\sqrt{\alpha_R^2+t^2}$, where $\alpha_R$ is the strength of Rashba spin-orbit coupling and $t$ is the strength of the interlayer hopping~\cite{Cavanagh2022}. Clearly, the odd-parity state is enhanced by the spin-orbit coupling and suppressed by the interlayer hopping. Under applied pressure, the spin-orbit coupling will remain more or less the same while interlayer hopping is expected to increase as the lattice constant decreases. Therefore, the even to odd-parity transition is naturally expected to be suppressed by pressure, illustrated in Fig.~4. The value of the relative strength $\widetilde{\alpha}$ (= $\alpha_R/t$) can not be obtained by fitting $H_{c2}-T$ data to equation (1), as both $T_c$ and $H_{orb}$ are fitting parameters in this equation whose $\widetilde{\alpha}$ dependence is not included. To get a qualitative estimation of $\widetilde{\alpha}$ under pressure, we compare our results to the bilayer model with Rashba spin-orbit coupling~\cite{Nogaki2022}. Our $H_{c2}-T$ diagrams closely resemble the $H-T$ phase diagrams calculated for different $\widetilde{\alpha}$ values (Fig. 2f and 2g), with pressure playing the same role as $\widetilde{\alpha}$. Based on the calculations, the even to odd-parity transition vanishes for $\widetilde{\alpha}$ values between 0 and 1. Accordingly, we estimate that $\widetilde{\alpha}$ is about 3 at abminet pressure and decreases to below 1 at a pressure around 3~GPa. 

\begin{figure}[ht!]
    \includegraphics[width=1\linewidth]{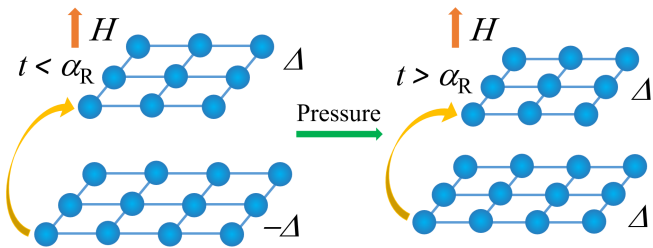}
    \caption{While Rashba-like spin-orbit coupling enables the odd-parity superconducting state in applied magnetic field, the interlayer hopping suppresses it. Under pressure, the strength of the interlayer hopping $t$ is expected to increase while the strength of the spin-orbit coupling $\alpha_R$ will remain roughly constant, leading to suppression of the odd-parity superconducting state.} 
\end{figure} 


Another possible mechanism for the parity change is the change in the nature of the quantum fluctuations. The relation between the quantum fluctuations and the superconductivity in CeRh$_2$As$_2$ has not been extensively studied since the superconducting phase diagram has been successfully explained in terms of the spin orbit coupling. However, our observation of two superconducting domes could suggest a different type of quantum fluctuations under pressure. At abmient pressure, CeRh$_2$As$_2$ is also close to a quantum critical point as evidenced by the normal state non-Fermi-liquid behavior, i.e., $C/T \sim T^{0.6}$ and $\rho (T) \sim \sqrt{T}$. This quantum critical point is proposed to be associated with a quadrupole order~\cite{Hafner2022}. Antiferromagnetic fluctuations have been observed from NMR measurements~\cite{kitagawa2022}. Theoretical calculations have shown that the critical field for the even to odd transition $H^\ast$ is enhanced by antiferromagnetic fluctuations, and $H^\ast/T_c$ stays nearly constant as $\widetilde{\alpha}$ varies~\cite{Nogaki2022}. Our results show that $H^\ast/T_c$ slightly decreases under pressure, then increases at 2.5~GPa. This seems to indicate a change of the strength in the antiferromagnetic fluctuations, in addition to the suppression of $\widetilde{\alpha}$. What this does to the parity of the superconducting state is yet to be explored theoretically.    

Under high pressure, valence fluctuations could develop in Ce-based compounds. A few other Ce-based superconductors also exhibit enhanced superconductivity under pressure~\cite{Holmes2004,Seyfarth2012,Jaccard1999,Miyake1999,Miyake2007,Ren2014}, which is attributed to the valence fluctuations associated with the valence crossover~\cite{Watanabe2011,Scheerer2018}. A method has been developed based on resistivity measurements to investigate the valence crossover and has been applied a wide range of Ce-based systems~\cite{Seyfarth2012}. It shows that when the system is tuned through the valence crossover by pressure, $\rho^{\ast} = \rho - \rho_0$ is strongly reduced by 1-2 order of magnitude, which is attributed to a sudden delocalization of 4$f$ electrons. We applied this analysis to CeRh$_2$As$_2$, as shown in Fig.~3. $\rho^{\ast}$ is more or less pressure independent in the whole pressure range without a clear indication of sudden drop crossing 2.5~GPa. A valence crossover is still possible at higher pressure, which will be investigated in future experiments.

To summarize, we observed a second dome of superconductivity in CeRh$_2$As$_2$ under pressure, which is dominated by an even-parity state in the whole magnetic field range. This parity change of superconductivity under pressure reveals the competition between the Rashba spin-orbit coupling and the interlayer hopping in determining the parity of superconducting states of systems with local symmetry breaking. The role of magnetic and valence fluctuations in determining the parity of the superconducting state requires future investigation. Our realization of the experimental tuning of the parity of the superconductivity provides insightful guidance for looking for new odd-parity superconductors.

We are highly indebted to I-lin Liu and Yuankan Fang for their continuous technique support to our pressure experiment. We would also like to thank Yoichi Yanase, Daniel Agterberg, Yifeng Yang and Li Yang for fruitful discussions.




\bibliography{CeRh2As2}

\clearpage

\end{document}